\documentstyle[12pt,a4,epsf]{article}

\begin{document}
\setlength{\topmargin}{-0.6in}
\setlength{\oddsidemargin}{0.2in}
\setlength{\evensidemargin}{0.5in}
\renewcommand{\thefootnote}{\fnsymbol{footnote}}

\begin{center}
{\Large {\bf D2-branes with magnetic flux in the presence
of RR fields}}
\vspace{0.8cm}

{D. K. Park,\raisebox{0.8ex}{\small a,b}\footnote[1]
{Email:dkpark@hep.kyungnam.ac.kr}
S. Tamaryan\raisebox{0.8ex}{\small c,d}\footnote[2]
{Email:sayat@moon.yerphi.am}
and H.J.W. M\"uller--Kirsten\raisebox{0.8ex}{\small c}\footnote[3]
{Email:mueller1@physik.uni-kl.de}}\\

\raisebox{0.8 ex}{\small a)}{\it Department of Physics,
Kyungnam University, Masan, 631-701, Korea}

\raisebox{0.8 ex}{\small b}{\it Michigan Center for Theoretical Physics,
Randall Laboratory, Department \\ 
of Physics,
University of Michigan, Ann Arbor, MI 48109-1120, USA}

\raisebox{0.8 ex}{\small c)}{\it Department of  Physics,
 University
of Kaiserslautern, 67653 Kaiserslautern, Germany}

\raisebox{0.8 ex}{\small d)}{\it  Theory Department, Yerevan
Physics Institute, Yerevan-36, 375036, Armenia}

\end{center}
\vspace{0.3cm}

{\centerline {\bf Abstract}}
\noindent
D2-branes  are studied in the context of Born-Infeld
theory as a source of the 3-form RR gauge potential.  Considering
the static case with only a radial magnetic field
it is shown that a locally stable hemispherical deformation
of the brane exists which minimises the energy locally.  Since
the D2-brane carries also the charge of D0-branes,
and the RR spacetime potential is unbounded from below, these can tunnel
to condense on the D2-brane.  The corresponding
instanton-like  configuration and the tunneling rate are
derived and discussed.

\vspace{2cm}
\centerline{PACS Numbers:11.10.Lm, 11.27.+d, 05.70.Fh}

\vspace{0.5cm}

\section{Introduction}

The low energy dynamics of D-branes
described by Born-Infeld theory on the worldvolume
of the brane is a topic of intense investigation
and has led to useful insights into
how string theory is interwoven with
electromagnetic phenomena. Such investigations  are
useful in domains where gravitational
(closed string) effects
can be ignored in the leading approximation. The dynamics
of these branes changes drastically in the
presence of external (spacetime) forces, such
as those of RR fields contained in type II
superstring theories, since the D-brane
carries the appropriate RR charges and
so couples to the potentials. In the following we will
be concerned with these
interactions in the particular case of D2-branes.

\vspace{0.3cm}
Born-Infeld theory by itself (i.e. without
RR fields) has been shown \cite{1} to imply
string-like brane excitations which
owing to their charge and tension
 can be identified in the appropriate limit
as fundamental strings. Such a string  may be looked at as  a
collapsed brane or the original brane as one with
a dissolved fundamental string \cite{2}. The conserved (axion)
charge of these strings along (say) $x^1$ results
from the electric component $B_{01}$ of the NS
B-field contained in the Born-Infeld action, and their tension
is of order 1 (as distinct from the $1/$(string coupling $g$)
behaviour of the tension of D-branes).
Considering Born-Infeld theory of the D2-brane
in the static limit and with only
the electric component of the
$U(1)$ gauge field in its world
volume, these strings are globally stable. However,
on application of the RR field, i.e.
with minimal coupling of the brane to the
RR potential, the
Born-Infeld string has been shown to be only
locally stable and can tunnel to or expand into
a D2-brane \cite{2,3}, which is also described
as the polarisation of a system of fundamental strings
into a higher dimensional brane. This tunneling
has been considered in detail in ref. \cite{3}.

\vspace{0.3cm}
It is natural to extend such investigations to the magnetic
counterpart or rather to the fully electromagnetic
formulation which, of course, introduces complications
as soon as Lorentz boosts or deviations from a  static
case are required.  Such investigations have been
carried out recently in various
directions \cite{4,5,6,7,8,9,10,11}, and our objective here is
to extend some considerations of refs. \cite{2} and \cite{4}
with particular reference to 
brane-antibrane pairs as in ref. \cite{1} and
the case of dielectric branes \cite{10,11}.

\vspace{0.3cm}
Since the D2-brane couples not only to the three-form
RR potential but in the presence of magnetic flux
also to the
one-form potential of D0-branes,
the magnetic case is very different from that of
the purely electric case.  Static torus-like
brane configurations in the presence of only a pure
magnetic field have been derived in ref.\cite{4} as well
as locally stable spherical configurations which can be
related to the dielectric D-branes of ref.\cite{10}.
The presence of the magnetic flux provides these
theories locally with energy minima at nontrivial
expanded configurations analogous to the
separation of charges in a dielectric medium.
The related stabilisation of branes (i.e. prevention
of their collapse to trivial or pointlike 
pure tension configurations)
by the presence of magnetic flux  was pointed out
earlier in ref.  \cite{12} in the context of WZW models.

\vspace{0.3cm}
In the following we remain in the context of the
model of refs.\cite{2} and \cite{4} 
and show that a locally stable
hemispherical brane configuration can be shown to
exist for a pure magnetic field in the
world volume. We demonstrate this
explicitly by considering fluctuations
around the  brane. The considerations are analogous
to those of branes and their antibranes of ref. \cite{1}
whose stability was studied in detail in ref. \cite{13}.
 We then consider the
Euclidean time pseudoparticle  brane configuration and its
relation to the classical brane  solution. Having found these, we
calculate the transition rate of the 2-brane
through the RR potential barrier and interpret
the result as a set of D0-branes condensing
on the D2-brane as suggested some time ago
\cite{12}. Irrespective of the physical
significance of the result, we consider the explicit
calculations which the model permits, to be
very instructive
for comparison with other cases
such as those of refs. \cite{1} and \cite{13},
and  in providing hints on what one
may expect in higher dimensional cases.
Some specific calculations, such as the comparison
with the spherical D2-brane of ref. \cite{4,10}, 
the evaluation of the fluctuation determinant 
and the Lorentz transformation of the action
are shifted into appendices.

\section{D2-branes in the presence of both electric and
magnetic fields}

The Born-Infeld action describing a $D2$-brane
as the source of the 3-form RR gauge potential $A_{\mu\nu\rho}$
 and its
coupling to this RR 3-form gauge potential
in type $IIA$ superstring theory
is
\begin{eqnarray}
I&=&-T_2\int d^3\xi\bigg\{\sqrt{-det\bigg(\partial_{\alpha}X^{\mu}
\partial_{\beta}X_{\mu}+2\pi \alpha^{\prime}{ F}_{\alpha\beta}\bigg)}\nonumber\\
&&+\frac{1}{3!}\epsilon^{\alpha\beta\gamma}A_{\mu\nu\rho}\partial_{\alpha}
X^{\mu}\partial_{\beta}X^{\nu}\partial_{\gamma}X^{\rho}\bigg\},
\label{1}
\end{eqnarray}
where $T_2 = 1/4\pi^2 g$ is the $p=2$ volume
tension of the $D2$--brane obtained from
\begin{equation}
T_p=\frac{2\pi}{(2\pi l_s)^{p+1}g}
\label{2}
\end{equation}
where we set $l_s=1$ and $\lambda=2\pi l_s^2=2\pi\alpha^{\prime}=2\pi$.
Here $\mu_p = T_p$ is the RR charge of the
brane under the $(p+1)-$form RR potential.
We follow ref.\cite{2} but include also a magnetic field.
  With $\alpha^{\prime} = 1$,
\begin{equation}
X^0=t, X^1=z, X^2=R(t,z)\cos\theta, X^3=R(t,z)\sin\theta, others = const.
\label{3}
\end{equation}
and
\begin{equation}
{ F}_{\theta z}=\partial_{\theta}A_z-\partial_zA_{\theta}
\label{4}
\end{equation}
and
\begin{equation}
H=dA, H_{0123}=h
\label{5}
\end{equation}
and
the target space metric with signature $-1,+1,+1,+1\cdots$,
and $ \alpha, \beta = t, z, \theta, E_z=2\pi{ F}_{tz},
E_{\theta} =2\pi{ F}_{t\theta}, B=2\pi{ F}_{\theta z} $:
\begin{eqnarray}
&&\bigg(\partial_{\alpha}X^{\mu}\partial_{\beta}X_{\mu}
+2\pi\alpha^{\prime}{ F}_{\alpha\beta}\bigg)\nonumber\\
&=&\left(\begin{array}{ccc}
-1+{\dot R}^2 & {\dot R}R^{\prime}+E_z & E_{\theta}\\
-E_z+R^{\prime}{\dot R} & 1+{R^{\prime}}^2 & B \\
-E_{\theta} &-B & R^2
\end{array}\right).
\label{6}
\end{eqnarray}
Then
\begin{equation}
I=\int dtdzd\theta {\cal L}(R,A_0, A_z, A_{\theta})
\label{7}
\end{equation}
where for $E_{\theta}=0$
\begin{equation}
{\cal L} = \frac{1}{4\pi^2g}\bigg\{-
\sqrt{R^2(1-{\dot R}^2+{R^{\prime}}^2-E^2_z)
+B^2(1-{\dot R}^2)}+\frac{h}{2}R^2\bigg\}.
\label{8}
\end{equation}
The four resulting Euler-Lagrange equations reduce to two
constraints and one equation of motion but imply also certain
conditions and therefore have to be considered carefully.
The equations are respectively: 
\begin{enumerate}
\item
For $R$ we obtain the equation of motion
\begin{eqnarray}
-&&\frac{\partial}{\partial t}\bigg(\frac{R^2{\dot R}+
{\dot R}B^2}
{\sqrt{R^2(1-{\dot R}^2+{R^{\prime}}^2-E^2_z)
+B^2(1-{\dot R}^2)}}\bigg)\nonumber\\
&&+\frac{\partial}{\partial z}
\bigg(\frac{R^{\prime}R^2}{\sqrt{R^2(1-{\dot R}^2+{R^{\prime}}^2-E^2_z)
+B^2(1-{\dot R}^2)}}\bigg)\nonumber\\
&&-\frac{R(1-{\dot R}^2+{R^{\prime}}^2 - E^2_z)}
{\sqrt{R^2(1-{\dot R}^2+{R^{\prime}}^2-E^2_z)
+B^2(1-{\dot R}^2)}}
+hR = 0.
\label{9}
\end{eqnarray}
\item
For $A_0$ we obtain the equation equivalent to the Gauss law
in Maxwell theory, i.e.
\begin{equation}
\frac{\partial D_E}{\partial z} = 0, \;\;
D_E\equiv
\frac{R^2 E_z}{\sqrt{R^2(1-{\dot R}^2+{R^{\prime}}^2-E^2_z)
+B^2(1-{\dot R}^2)}}.
\label{10}
\end{equation}
\item
For $A_{\theta}$ we obtain the equation
\begin{equation}
\frac{\partial (2\pi D)}{\partial z} = 0, \;\;
2\pi D\equiv\frac{2\pi B(1-{\dot R}^2)}
{\sqrt{R^2(1-{\dot R}^2+{R^{\prime}}^2-E^2_z)
+B^2(1-{\dot R}^2)}}.
\label{11}
\end{equation}
\item
Finally for $A_z$ we obtain the equation
\begin{equation}
\frac{\partial}{\partial t}( 2\pi D_E) +\frac{\partial}{\partial \theta}
(2\pi D) = 0.
\label{12}
\end{equation}

\end{enumerate}
We observe that for the purely electric case considered in ref. \cite{2}
the second and the fourth equations imply that the electric 
 quantity $D_E$
is independent of both $z$ and $t$ and is therefore a constant. 
In the purely magnetic case, which we concentrate on here, we
conclude from the third and fourth equations
 that the magnetic quantity $D$ does not possess an
explicit  $z$-dependence, nor an
explicit  $\theta$-dependence, but can depend explicitly on $t$.
In addition - and this is a vital point -
$D$ is a functional
of $R$ which is a function of $z$ and $\theta$ and possibly of $t$.

In the static and purely magnetic case the two remaining
equations are
\begin{equation}
\frac{\partial}{\partial z}
\bigg(\frac{R^{\prime}R^2}{\sqrt{R^2(1+{R^{\prime}}^2)
+B^2}}
\bigg)
-\frac{R(1+{R^{\prime}}^2 )}
{\sqrt{R^2(1+{R^{\prime}}^2
+B^2)}}
+hR = 0.
\label{13}
\end{equation}
and
\begin{equation}
 D = \frac{2\pi B}{\sqrt{R^2(1+{R^{\prime}}^2
+B^2)}}
\label{14}
\end{equation}
Setting 
$$
P^2=1/(1-D^2)
$$ 
and solving the latter equation for $B$
we obtain
\begin{equation}
B=PDR\sqrt{1+{R^{\prime}}^2}
\label{15}
\end{equation}
and 
\begin{equation}
R^2(1+{R^{\prime}}^2)+B^2=R^2(1+{R^{\prime}}^2)P^2
\label{16}
\end{equation}
This equation allows us to express the quantity $P^2$
and so  $D$ in terms of
$R$ and $B^2$. Using this,  eq.(\ref{13})
leads to
\begin{equation}
\frac{\partial}{\partial z}\bigg(\frac{RR^{\prime}}{P\sqrt{1+{R^{\prime}}^2}}
\bigg)-\frac{1}{P}\sqrt{1+{R^{\prime}}^2} +hR = 0.
\label{17}
\end{equation}
With further  manipulations this equation can be converted into
\begin{eqnarray}
&&
\frac{d}{dz}\bigg(\frac{R}{P\sqrt{1+{R^{\prime}}^2}}-\frac{hR^2}{2}\bigg) = 0,
\nonumber\\
&&\frac{R}{P\sqrt{1+{R^{\prime}}^2}}-\frac{hR^2}{2}=C,
\label{18}
\end{eqnarray}
where $C$ is a constant.
Eq.(\ref{18}) is also contained in the work of ref.\cite{4}. 
and can also be obtained from a Legendre transformed 
Lagrangian density
\begin{equation}
{\cal L}_B:=-{\cal L} +B\frac{\partial{\cal L}}{\partial B}, \;\;
D=-4\pi^2g\frac{\partial{\cal L}}{\partial B}
\label{19}
\end{equation}
where
\begin{equation}
{\cal L}_B(R)=\frac{1}{4\pi^2g}\bigg\{\frac{R\sqrt{1+{R^{\prime}}^2}}{P}
-\frac{h}{2}R^2\bigg\}
\label{20}
\end{equation}
We comment on other derivations later.

We now define the Hamiltonian density ${\cal H}$ by
\begin{equation}
{\cal H} = P_R{\dot R} + P_{A_z}{\dot A}_z+ P_{A_{\theta}}{\dot A}_{\theta}
-{\cal L},
\label{21}
\end{equation}
where $P_R, P_{A_z}, P_{A_{\theta}}$ are the conjugate momenta
of $R, A_z, A_{\theta}$, i.e. $P=\partial{\cal L}/\partial{\dot R}$, etc.
Then
\begin{equation}
{\cal H} =\frac{1}{4\pi^2 g}\bigg\{
\frac{ R^2+B^2
+R^2{R^{\prime}}^2}{\sqrt{(1-{\dot R}^2)( R^2+B^2)
+R^2({R^{\prime}}^2 -E^2_z)}}-\frac{h}{2}R^2\bigg\}+\frac{1}{g}D_EA^{\prime}_0
\label{22}
\end{equation}
This expression is to be supplemented by the 
 two constraint equations (\ref{10}) and (\ref{11})
for $E_z$ and $B$.

The magnetic flux in the world volume is for ${\dot R}=0$ given by
\begin{equation}
N=\frac{1}{2\pi}\int d\theta dz{ F}_{\theta z}
=\frac{1}{2\pi}\int dz\frac{R^2D\sqrt{1+{R^{\prime}}^2}}
{\sqrt{D^2_E+R^2(1-D^2)}}.
\label{23}
\end{equation}
This integral of the worldvolume 2-form field strength is
in general  not finite.  Thus to obtain a finite
expression for this charge, one has to close the membrane
configuration  as already discussed in ref. \cite{14}.
This is essentially the charge associated with the
worldvolume vector potential. The latter is a 1-form
which couples to a $D0$-brane.

The Hamiltonian $H$ now becomes in going to the static case
(${\dot R}=0$)
\begin{equation}
H=\frac{1}{4\pi^2 g}\int dz d\theta \bigg[(R^2+D^2_E)\sqrt{\frac{(1+{R^{\prime}}^2)}
{(D^2_E+R^2-R^2D^2)}}-\frac{h}{2}R^2\bigg].
\label{24}
\end{equation}
In performing the variation
$$
\frac{d}{dz}\bigg(\frac{\delta{H}}{\delta R^{\prime}}\bigg)
-\frac{\delta {H}}{\delta R} = 0
$$
one has to remember that the expression $D$
is a functional of $R$ as expressed by
eq.(\ref{16}). Considering the static magnetic case
 this means that before the variation of the Hamiltonian
\begin{equation}
H=\frac{1}{4\pi^2 g}\int dz d\theta \bigg[PR\sqrt{1+{R^{\prime}}^2}
-\frac{h}{2}R^2\bigg]
\label{25}
\end{equation}
is performed, one has to replace $P$ with the help of 
 expression (\ref{16}) in terms of $B$.
Finally again replacing the expression with $B$ by that with $P$
one arrives at the same equation which was obtained 
earlier from the Lagrangian. 
The Hamiltonian will be needed later for the calculation of
the energy of our minimum energy configuration.
In the above we considered the case of a cylindrical geometry.
For comparison we provide in Appendix A
the main formulae (for flux and energy)
of  the spherical case considered in
refs. \cite{4,10}.

\section{Pure magnetic case}

We consider the static and purely  magnetic case.
 The total magnetic flux is nonzero which means
 that we have objects dissolved in the D2-brane carrying
 magnetic charge. Being the singularities of the magnetic field,
 these objects must be pointlike, hence they are D0-branes. In
ref. \cite{4}
it was shown that the  number of D0-branes
can in this case be simply related to the
 magnetic flux.
The
Hamiltonian $H$ is given by eq. (\ref{24}).
We know that if the $D2$ brane is coupled to  magnetic flux
it also carries the charge of the D0 brane \cite{15}.
Thus it is sensible to consider the limit of the
spherical configuration shrinking to a point.
The worldvolume magnetic flux is given by
\begin{equation}
N=\frac{PD}{2\pi}\int dz R\sqrt{1+R^{\prime 2}}=\frac{PDS}{(2\pi)^2},\;\;
S=\int dz d\theta R\sqrt{1+{R^{\prime}}^2},
\label{26}
\end{equation}
where $S$ is the area of the world volume which
follows simply from the fieldless static metric contained 
in eq. (\ref{6}). One can see from this relation that if
$D$ were kept  constant in a variation, a nonvanishing variation of $S$
would have to be compensated by a nonvanishing variation of
$N$, and thus the number of $D0$-branes would not be fixed
unless compensated appropriately from a reservoir of
D0-charge around the D2-brane.
From the relation (\ref{26}) we obtain
\begin{equation}
D^2=\frac{(NT_0)^2}{(ST_2)^2+(NT_0)^2},\;\; NT_0=ST_2DP,
\label{27}
\end{equation}
so that for $S$ shrinking to zero $D\rightarrow 1$ and
the energy can be presented as the mass  of $N$ D0-branes, i.e.
\begin{equation}
E_{R\rightarrow 0} = T_0N.
 \label{28}
\end{equation}
One can  also express the energy like eq.(\ref{27}), as observed
earlier in ref.\cite{4}:
\begin{equation}
H=\sqrt{(ST_2)^2+(NT_0)^2}-\frac{h}{4\pi g}\int dz R^2.
\label{29}
\end{equation}
Varying this expression (and so the surface area $S$ with respect to $R$)
one again obtains the same equation of motion as with the other
 methods for $D=D(R)$.
 The equation  of motion, eq.(\ref{18}),
can be rewritten to give
\begin{equation}
R^{\prime}=\pm\frac{h}{hR^2+2C}\sqrt{(R^2_+-R^2)(R^2-R^2_-)}
\label{30}
\end{equation}
with
\begin{equation}
R^2_+R^2_-=\frac{4C^2}{h^2},\qquad 
R^2_++R^2_-=\frac{4}{P^2h^2}(1-CP^2h).
\label{31}
\end{equation}



We are looking for nonperiodic, finite energy solutions. Consequently we
set $R_-=0$ yielding for the integration constant $C=0$.
Then eq.(\ref{30}) simplifies to
\begin{equation}
 R^{\prime}=\pm\frac{\sqrt{R^2_+-R^2}}{R},
\label{32}
\end{equation}
where $R_+=\frac{2}{Ph}$ and the configuration
$R$ is geometrically a radius. The solutions  of this equation are
the configurations $R$ with respectively positive  ($z_0>z$) 
and negative ($z_0<z$) derivatives as in eq. (\ref{32}), i.e.
\begin{eqnarray}
R(z)&=&-\int^z_{z_0-R_+}\frac{(z-z_0)dz}{\sqrt{R^2_+-(z-z_0)^2}}
=\sqrt{R^2_+-(z-z_0)^2},\nonumber\\
R(z)&=&+\int^{z_0+R_+}_z\frac{(z-z_0)dz}{\sqrt{R^2_+-(z-z_0)^2}}
=\sqrt{R^2_+-(z-z_0)^2}
\label{33}
\end{eqnarray}
with the enveloping  sphere
\begin{equation}
R^2+(z-z_0)^2=R^2_+.
\label{34}
\end{equation}
This sphere is thus  the envelope of  the pair of 
spherical shells with $z_0-R_+\leq z \leq z_0, z_0\leq z
\leq z_0+R_+$, or the pair of circles of radius $R$
at positions
\begin{eqnarray}
z&=&z_0 \pm \sqrt{R^2_+ - R^2}\nonumber\\
&=&\pm \int^R_{R_+}\frac{R dR}{\sqrt{R^2_+ - R^2}}.
\label{35}
\end{eqnarray}
Thus $z(R)$ is double valued.  The two
possible signs define the two hemispherical  configurations on
the enveloping sphere given by eq. (\ref{34})
and as indicated in Fig.1. 
These
two hemispherical 
configurations defined by respectively positive or negative
$R^{\prime}$ (observe that $R$ originally defined
as a radius is always
positive, the variable angle $\theta$
being understood)  can be looked at as a brane and its antibrane analogous to
the appearance of branes and their antibranes in ref.\cite{1}.
In fact, that the sphere  (\ref{34}) represents an unstable
 brane-antibrane
pair can be seen by differentiating (\ref{34}) with respect to $z$
and re-inserting the equation, which gives
\begin{equation}
\frac{dR}{dz} = \frac{z-z_0}{\sqrt{R^2_+-(z-z_0)^2}}.
\label{36}
\end{equation}
We see that this is an odd function which reverses its sign
on passing through $z_0$. This is one of the characteristic
properties  of the configuration called a bounce \cite{16}.
(Like a periodic instanton it can be loosely looked at
as an instanton-antiinstanton pair, the instanton (or antiinstanton)
being a monotonically increasing (or decreasing) function
of its argument contrary to the behaviour of the combination).
The sphere is the limiting form of a spheroidal
bulge like that discussed in the electric case in ref. \cite{2};
hence the behaviour of $R^{\prime}$ here is
also that in this limit, i. e. that  of a
function with the shape typical of an odd  first excited state
wave function.
Thus as expected in ref. \cite{1} for the D3-brane model
considered there  and demonstrated for this in
ref. \cite{13}, the brane-antibrane pair is unstable, which means that
small fluctuations in its neighbourhood possess a negative
mode, the tachyon (how the brane-antibrane system with
magnetic flux $N=2$  is related 
to tachyon condensation is described for instance
 in ref. \cite{15}).
The appearance of these brane-antibrane
pairs is a consequence of the two possible signs of the
derivative in eq. (\ref{32}), and
so of two possible solutions,
and these in turn are a consequence
of the square-root form of the Born-Infeld Lagrangian
density.  Thus Born-Infeld theory leads very naturally
to brane-antibrane configurations through a linkage of
the solutions or continuation of the one solution
to the other  with the opposite sign of
its derivative as in the cases considered in refs. \cite{1} and
\cite{2}.

For the energy $E$  of the solution  with
$R\sqrt{1+{R^{\prime}}^2} = R_+$ 
and $z_0-R_+\leq z \leq z_0 $ as integration
domain
 one finds from $H$
\begin{equation}
E=\frac{NT_0}{2D} -\bigg(\frac{4\pi R^3_+}{3}\bigg)\frac{ hT_2}{2}.
\label{37}
\end{equation}
With eq.(\ref{27}) the first part
can be rewritten so that
\begin{equation}
E=\frac{1}{2}\sqrt{(NT_0)^2+(ST_2)^2}-
\bigg(\frac{4\pi R^3_+}{3}\bigg)\frac{hT_2}{2}
\label{38}
\end{equation}
(With the factor 1/2 the volume part is the same as
that in the electric case of ref. \cite{2}).
This result is very physical with the first part
representing the square root of the sum of
the squares of the masses of the D2-brane and
the N D0-branes, and the
second the energy stored in the volume.
The negative sign of the latter  indicates that the
(hemi)spherical configuration 
of radius $R_+$ minimises the energy locally as will be shown below.
We observe from eqs.(\ref{24}) and (\ref{25}) that the
potential is unbounded below for large values of $R$.
Thus tunneling of the locally stable hemispherical configuration
is possible,  and one can calculate
the appropriate transition rate. This will be 
examined in the following.

\section{Small fluctuations and the pre-exponential factor}

 We now  consider the second variation of
$H$ around  the positive derivative  solution (\ref{33})
 and demonstrate that it is positive definite
 under small fluctuations. This means that any small
 deviation from this  solution of (\ref{33})
 increases the energy. Consequently we have a minimum
of the energy functional. This minimum is not
 a global minimum because the potential is unbounded below  as observed
earlier; thus it is only  a local minimum.
 Next we determine the fluctuation operator describing the behaviour
 of the second variation of the energy functional in the vicinity of
 this  classical solution.
 Born-Infeld  theory is  a covariant theory; therefore the existence of the
static finite energy solution
 implies  the existence of the instanton-type pseudoparticle  solution
which is really a bounce.
 In calculating the tunneling transition
rate in the next section,
we  use the semiclassical approximation  around this instanton-type  solution.
 The argument of the exponential  of the semiclassical amplitude
is the pseudoparticle  action, and the  pre-exponential factor
 is, as usual,
 the determinant of the fluctuation operator of the 
pseudoparticle solution.
 From  Lorentz invariance
it follows that the
fluctuation operators of both the static and the
 pseudoparticle  solutions must coincide up to a possible
sign factor, so that it suffices  to calculate
 the determinant in the static case.

The second variation of $H$ is given by
\begin{equation}
\delta^2H=\frac{1}{2}\int dz\bigg[
\frac{{\delta}^2H}{{\delta}{R^{\prime}}^2}
(\delta R^{\prime})^2+
2\frac{{\delta}^2H}{{\delta R}{\delta {R}^{\prime}}}
\delta R\delta R^{\prime}+
\frac{{\delta}^2H}{{\delta R}^2}(\delta R)^2\bigg]
\label{39}
\end{equation}
This is evaluated in Appendix B and the result is
\begin{equation}
{\delta}^2H=\frac{1}{2}\int dz\delta R{\hat M}\delta R
=\frac{1}{2}\int \frac{dR}{R^{\prime}}\delta R{\hat M}\delta R
\label{40}
\end{equation}
where the fluctuation operator ${\hat M}$ is
\begin{equation}
{\hat M}=-\frac{1}{2\pi gP}\frac{1}{R^{\prime}}\frac{d}{dz}
\frac{R{R^{\prime}}^2
(1+D^2{R^{\prime}}^2)}
{(1+{R^{\prime}}^2)^{3/2}}\frac{d}{dz}\frac{1}{R^{\prime}}.
\label{41}
\end{equation}
The derivative $R^{\prime}$ is positive for one hemisphere
as explained earlier, and negative for the other. Thus
one of these minimises the energy locally, the other does not.
We present  ${\hat M}$ as a product because this  has the
 advantage that its determinant can be easily calculated.
This is our next step.

The determinant of the fluctuation operator ${\hat M}$ must be normalized.
 As a
normalization point we choose $ D=0  (P=1)$ which is  the case of
a vanishing
magnetic field.
We denote  the corresponding classical solution by $R_0$
 and  the corresponding fluctuation operator by ${\hat M_0}$.
The normalized determinant is then
\begin{equation}
det_n{\hat M}=\frac{det{\hat M}}{det{\hat M_0}}.
\label{42}
\end{equation}
We use the following two  properties of  determinants:
\begin{equation}
det(AB)=det(BA),\qquad \frac{det(TA)}{det(TB)}=\frac{detA}{detB}.
\label{43}
\end{equation}
The expression for the normalized determinant then simplifies and we have
\begin{equation}
det_n{\hat M}=\frac{det[R(1+D^2{R^{\prime}}^2)(1+{R^{\prime}}^2)^{-3/2}/P]}
{det[R_0(1+{R^{\prime}}^2_0)^{-3/2}]}.
\label{44}
\end{equation}
For our solution with  integration constant $C=0$ we have
\begin{equation}
\frac{R(1+{R^{\prime}}^2)^{-3/2}}{P}=\frac{h^3 R^4}{8}P^2, \;\;
(1+D^2{R^{\prime}}^2)=\frac{1}{P^2}+\frac{D^2R^2_+}{R^2}
\label{45}
\end{equation}
resulting in
\begin{equation}
det_n{\hat M}=\frac{det(R^4 +R^2P^2D^2R^2_+)}{detR^4_0}.
\label{46}
\end{equation}
The expression $ R^4 +R^2P^2D^2R^2_+ $ is a c-number
and its determinant yields an integral:
\begin{eqnarray}
det(R^4 +R^2P^2D^2R^2_+ )&=&exp[Tr[\ln (R^4 +R^2P^2D^2R^2_+ )]]\nonumber\\
&=&exp[\int ^{0}_{-R_+} \frac{dz}{R_+}\ln ( R^4 +R^2P^2D^2R^2_+)].
\label{47}
\end{eqnarray}
The evaluation  of the integral is straightforward and yields
\begin{equation}
\int^{0}_{-R_+}
 \frac{dz}{R_+}\ln (R^4 +R^2P^2D^2R^2_+ )=
-4+2\ln(2R^2_+)+\ln(P^2-1)+P\ln\bigg(\frac{P+1}{P-1}\bigg)
\label{48}
\end{equation}
which gives for the determinant
\begin{equation}
det_n{\hat M}=\frac{P^2-1}{P^4}\bigg(\frac{P+1}{P-1}\bigg)^P
=\frac{D^2}{P^2}\bigg(\frac{P+1}{P-1}\bigg)^P.
\label{49}
\end{equation}
One may note that the expressions (\ref{48})  and (\ref{49})
are finite for $P \rightarrow 1$.

\section{The tunneling amplitude}

To describe the tunneling by the  instanton-type
pseudoparticle  solution in Euclidean
time, we make a Wick  rotation and set $t\rightarrow i \tau$.
 The way   to find this pseudoparticle  solution in  a  case
like  the one here was  already proposed in \cite{2}.
 One performs  an exchange $\tau \leftrightarrow z$,
 so that  having the static solution one obtains
  the  pseudoparticle  one.
However, the Lagrangian of  eq.(\ref{8}) does  not exhibit
 an  explicit symmetry under this   exchange.
This might seem strange but has a simple explanation
if we include in our considerations
 the polar component $E_{\theta}$ of the electric field
which has therefore been given explicitly
 in eq.(\ref{6}).   We now perform  a  right angle rotation in the
 $(z,\tau)$ plane, i.e.
\begin{equation}
z\rightarrow z^{\prime}=\tau, \qquad \tau\rightarrow  {\tau}^{\prime}=-z
\label{50}
\end{equation}
under which electromagnetic components transform as follows:
\begin{equation}
E_z  \rightarrow  E_z
\label{51}
\end{equation}
and
\begin{equation}
E_{\theta}  \rightarrow  B, \;\;
B  \rightarrow  -E_{\theta},
\label{52}
\end{equation}
where B is as before the magnetic field.
The above transformations show that in the purely
 magnetic case the action does  not have the  explicit
symmetry under the exchange $\tau \leftrightarrow z$.
On the other hand the  transformations (\ref{52})
are electromagnetic duality transformations \cite{17}
leaving the equations of motion invariant \cite{17,18}.
Therefore the Lorentz
transformations yield for the pseudoparticle  the same equation (\ref{26}).
We verify this explicitly in Appendix C.
Also one can make the backward  Lorentz rotation and obtain
 the pseudoparticle solution
in the original frame.
 The shape of the pseudoparticle is that of a sphere which,
however,   is not invariant
under this rotation and instead becomes  an ellipsoid.
 It is instructive  to obtain the
same ellipsoid directly from the Euler-Lagrange  eq.(\ref{9}).
We set $E_z=0, R^{\prime}=0$, so that in this nonstatic case
$$
B^2=\frac{D^2R^2}{1-{\dot R}^2-D^2}
$$ 
and obtain for the time dependent solution  the
equation
\begin{equation}
-\frac{d}{dt}\bigg(\frac{R \dot R}{\sqrt{1-{\dot R}^2-D^2}}\bigg)
-\sqrt{1-{\dot R}^2-D^2}+hR=0
\label{53}
\end{equation}
Going  to  Euclidean time with  $t=i\tau $
 and using  the fact that the equation
does not contain $\tau $ explicitly,
 allows us to convert it into the following
first order differential equation
\begin{equation}
\frac{R}{P^2\sqrt{1-D^2+(\frac{dR}{d\tau})^2}}-\frac{h}{2}R^2=C,
\label{54}
\end{equation}
where $C$ is a constant. For the discussion below
it is more transparent to consider first the general case
with $C\neq 0$. Then
\begin{equation}
\frac{dR}{d\tau}=\pm\frac{h}{2P(C+\frac{h}{2}R^2)}
\sqrt{(R^2_+-R^2)(R^2-R^2_-)}
\label{55}
\end{equation}
where
\begin{equation}
R^2_{\pm}=\frac{2}{P^2h^2}\bigg[1-ChP^2\pm\sqrt{1-2ChP^2}\bigg]
\label{56}
\end{equation}
For large $R$ the energy given by the Hamiltonian $H$ of eq.(\ref{25})
decreases without limit. Thus the motion of
 the pseudoparticle in 
Euclidean time starts with zero `` velocity'' $dR/d\tau$
at $R=R_+$ and bounces back from the potential
wall at point $R_-$ at  the time-symmetric point
$\tau = \tau_0$ until it again reaches $R=R_+$. At
the time-symmetric point the reversal of the velocity
implies a switch from one sign of the square root
to the other.     
Now we again  choose the integration constant $C$ to be zero as in eq.
(\ref{32}). We thus have $R_- =0$ and obtain
\begin{equation}
 P^2\bigg(\frac{dR}{d\tau}\bigg)^2 + 1 = \frac{R^2_+}{R^2},
\label{57}
\end{equation}
From this we obtain the  solution
\begin{equation}
P^2\bigg(\sqrt{R^2-R^2_+}-\sqrt{R^2_--R^2_+}\bigg)^2
+(\tau-\tau_0)^2=0,
\label{58}
\end{equation}
where $R_-$ is the bouncing or turning  point at Euclidean time
$\tau_0$. 
However, in evaluating the action of the bounce
 and hence the tunneling rate
it is more convenient  to integrate with respect
to $R$ by setting $d\tau = dR/{\dot R}$ and  using  the derivative relation
(\ref{57}), i.e. also
\begin{equation}
PR\sqrt{1+\bigg(\frac{dR}{d\tau}\bigg)^2}=\sqrt{R^2_++D^2P^2R^2}.
\label{59}
\end{equation}
The pseudoparticle  action $I_E$ is now obtained from eq.(\ref{7}), 
i.e.
\begin{equation}
I_E=-T_2\int dz\int d\tau\bigg\{-PR\sqrt{1+\bigg(\frac{dR}{d\tau}\bigg)^2}
+\frac{h}{2}R^2\bigg\}
\label{60}
\end{equation}
The orbit of the bounce is that from $R_+$ to $R_-$ and back,
so that
\begin{equation}
I_E=-2\pi PT_2\int dz\int^{R^2_+}_0 dR^2\bigg\{\sqrt{\frac
{R^2_++D^2P^2R^2}{R^2_+-R^2}} -\frac{h}{2}
\frac{R^2}{\sqrt{R^2_+-R^2}}\bigg\}
\label{61}
\end{equation}
Evaluating this by integrating over the compact length $L$
of a torus we obtain \cite{19}
\begin{equation}
I_E=-4\pi PT_2LR^2_+{_2F}_1(1,-1/2,3/2;-D^2P^2)
 +\bigg(\frac{4\pi R^3_+}{3}\bigg)LPhT_2
\label{62}
\end{equation}
Using eq.(\ref{27}) the action $I_E$ can be rewritten as
\begin{eqnarray}
I_E&=&-L\frac{NT_0}{D}{_2F}_1(1,-1/2,3/2;-D^2P^2)
 + \bigg(\frac{4\pi R^3_+}{3}\bigg) LPhT_2\nonumber\\
&=& -L\sqrt{(ST_2)^2+(NT_0)^2}{_2F}_1(1,-1/2,3/2;-D^2P^2)\nonumber\\
&&+L\bigg(\frac{4\pi R^3_+}{3}\bigg) PhT_2
\label{63}
\end{eqnarray}
Since action corresponds to energy $\times$ length, this
result is, as expected, similar
to the energy (\ref{38}) of the sphere but takes
into account the ellipsoidal deformation.
In computing the decay or tunneling 
rate we have to subtract from the action 
of the pseudoparticle the action of the initial state
which means here the square-root part of the action.
For the argument of the hypergeometric
function we also have
\begin{equation}
D^2P^2=\bigg(\frac{\pi N}{R^2_+}\bigg)^2.
\label{64}
\end{equation}
We observe that for no magnetic field and so  charge, or
no D0-branes,   the
radius $R_+$ is zero, that is, there is no dielectric effect.

With this  the semiclassical approximation
of the tunneling amplitude $\Gamma$ is given by
\begin{equation}
\Gamma=(det{\hat M})^{-\frac{1}{2}}exp(-I_E).
\label{65}
\end{equation}
Inserting the appropriate quantities,
 the final expression for the tunneling rate  is seen to be
\begin{eqnarray}
\Gamma&=&\frac{P}{D}\bigg(\frac{P-1}{P+1}\bigg)^{P/2}
exp\bigg(-L\frac{4\pi R^3_+}{3}PhT_2\bigg)\nonumber\\
&=& \frac{P}{D}\bigg(\frac{P-1}{P+1}\bigg)^{P/2}
exp\bigg(-\frac{8L}{3\pi gP^2h^2}\bigg).
\label{66}
\end{eqnarray}
In the limit of $D \rightarrow 0, P \rightarrow 1$,
 there is no electromagnetic
field (cf. (\ref{11}) and (\ref{16})). But we still  have a static
 spherical  solution which is the excitation
of the original brane under the
 influence of the RR field. In the opposite limit of
$D \rightarrow 1, P \rightarrow\infty  $ of
a  strong magnetic field,  it is better
to express the tunneling amplitude via the  D0-brane number. 
From eqs.(\ref{26}) and (\ref{27}) it
follows that
\begin{equation}
P^2=\bigg(\frac{4}{\pi h^2N}\bigg)^2\bigg[1+O\bigg(\frac{1}{P^2}\bigg)\bigg],
 \;\; \pi h^2N<<1.
\label{67}
\end{equation}
Thus the tunneling rate becomes
\begin{equation}
\Gamma \simeq  \frac{4}{\pi h^2 N} 
exp\bigg(-\frac{\pi L  h^2 N^2}{6g}\bigg).
\label{68}
\end{equation}

\section{The purely temporal case}

We consider here the cylindrical D2-brane with magnetic
field in a purely time-dependent case
and demonstrate an analogy with the static, purely electric case \cite{2,3}.
This permits the consideration of quantum classical
transitions in the magnetic  case to be taken over
from  the electric  case.

We 
start from our  original D2-brane action
\begin{equation}
S_{D_2}[R, A_{\theta}, A_z]=-T_2 \int dt d\theta dz
\left[\sqrt{R^2(1 + R'^2 -\dot{R}^2)+B^2
	     (1-\dot{R}^2)}-\frac{h}{2}R^2\right]
\label{69}
\end{equation}
with $-\pi \leq \theta \leq \pi$ and $-L/2 \leq z \leq L/2$.
Varying the action (\ref{69}) yields the following equations of motion:
\begin{eqnarray}
\frac{\partial}{\partial z} D  =
 \frac{\partial}{\partial \theta} D& =& 0, \nonumber\\  
\frac{\partial}{\partial t} \frac{(R^2 + B^2) \dot{R}}
  {\sqrt{R^2 (1 + R'^2 - \dot{R}^2) + B^2(1-\dot{R}^2)}}
&-&\frac{\partial}{\partial z} \frac{R^2 R'}
  {\sqrt{R^2 (1 + R'^2 - \dot{R}^2) + B^2 
			      (1-\dot{R}^2)}} \nonumber\\
+ \frac{R (1 + R'^2 - \dot{R}^2)}
{\sqrt{R^2 (1 + R'^2 - \dot{R}^2) + B^2(1-\dot{R}^2)}}
&-& h R = 0,
\label{70}
\end{eqnarray}
where now with $B=2\pi F_{\theta z}$
\begin{equation}
D = \frac{B (1 - \dot{R}^2)}
  {\sqrt{R^2 (1 + R'^2 - \dot{R}^2) + B^2
			     (1 -  \dot{R}^2)}}.
\label{71}
\end{equation}
Since  the equations $\partial_z D = \partial_{\theta}
 D = 0$ have  already been solved in the 
static case ($R = R(z)$), we  now consider the solution of 
$\partial_z D = \partial_{\theta} D = 0$ in the
 purely  temporal case, i.e. for  $R=R(t)$.

In this case  
$F_{\theta z} = F_{\theta z}(t)$ 
could be an arbitrary function of time.
Since, however, we are considering the purely
 magnetic case, it is appropriate
to assume again $F_{\theta z} = const$, because if not, the time-dependent 
magnetic field usually generates an electric field. Thus, we choose
\begin{equation}
F_{\theta z} = \frac{N}{L},
\label{72}
\end{equation}
where
\begin{equation}
\label{73}
N = \frac{1}{2 \pi} \int d\theta dz F_{\theta z}
\end{equation}
is the total number of D0-brane particles
or quantised flux through the cylindrical surface of length $L$.

Using (\ref{72}) and $R' = 0$  the last of  eqs.
(\ref{70})  reduces to 
\begin{equation}
\frac{\partial}{\partial t}
\left( \dot{R} \sqrt{\frac{R^2 + \xi^2}{1 - \dot{R}^2}}  \right)
+ R \sqrt{\frac{1 - \dot{R}^2}{R^2 + \xi^2}} - h R = 0
\label{74}
\end{equation}
where
\begin{equation}
\label{75}
\xi = \frac{N \lambda}{L}, \;\;\lambda =2\pi.
\end{equation}
One can also show that eq.(\ref{74}) is obtained  directly
 by varying the action
\begin{equation}
\label{76}
\tilde{S}_{D_2}[R] = -2\pi L T_2
\int dt \left[ \sqrt{(R^2 + \xi^2)(1 - \dot{R}^2)} - \frac{h}{2} R^2 \right].
\end{equation}

 In general, we cannot insert a  classical solution into the action
before varying it, however here
this is permissible for the constant solution (\ref{72}), which can
also be shown  by varying action (\ref{76}). 
Before solving (\ref{74}) it is helpful to consider the potential 
$V_{D_2}(R)$ which can be read off from (\ref{76}),
\begin{equation}
\label{77}
V_{D_2}(R) = 2 \pi L T_2
\left(\sqrt{R^2 + \xi^2} - \frac{h}{2} R^2 \right).
\end{equation}
It is interesting to reexpress the potential as
\begin{equation}
\label{78}
V_{D_2}(R) = \sqrt{(S T_2)^2 + (N T_0)^2} - h {\cal V} T_2
\end{equation}
where $S=2\pi R L$ and ${\cal V} = \pi R^2 L$. In fact,
 $S$ and ${\cal V}$ are
respectively
surface area and volume of the cylindrical D2-brane in flat spacetime. Hence, 
the potential consists of two terms,  i.e. the surface energy of the D2-brane
with $N$ D0-branes dissolved in it  and the  volume energy.

The shape of the potential $V_{D_2}(R)$ is as follows. If $\xi h > 1$, 
$V_{D_2}(R)$ is a
monotonically
 decreasing function and $R=0$ becomes a
point of instability. If $\xi h < 1$, $V_{D_2}(R)$ has 
a local  minimum at $R=0$ and 
a global  maximum at $R \equiv R_{\ast} = \sqrt{1/h^2 - \xi^2}$.
 Thus  we have
 quantum tunneling in this case. Here we  confine ourselves to  the
latter case ($\xi h < 1$). We  summarize  several particular values computed
from the potential:
\begin{eqnarray}
\label{79}
V_{D_2}(R=0)&=& 2\pi L T_2 \xi \equiv N T_0,   \\   \nonumber
V_{D_2}(R=R_{\ast})&=& \frac{\pi L T_2}{h} (1 + h^2 \xi^2),   \\  \nonumber
V_{D_2}^{''}(R=0)&=& \frac{1}{\xi} - h > 0,                   \\  \nonumber
V_{D_2}^{''}(R=R_{\ast})&=& - h (1 - h^2 \xi^2) < 0.
\end{eqnarray}

We  solve eq.(\ref{74}) for  $\xi h < 1$. 
Since we have  tunneling in this case, it is more convenient to go to
Euclidean time by introducing $\tau = - i t$. Then  eq.(\ref{74})
 becomes
\begin{equation}
\label{80}
-\frac{d}{d\tau}
\left( \dot{R} \sqrt{\frac{R^2 + \xi^2}{1 + \dot{R}^2}}  \right)
+ R \sqrt{\frac{1 + \dot{R}^2}{R^2 + \xi^2}} - h R = 0
\end{equation}
where a dot denotes   differentiation with respect to $\tau$. 
In fact, eq.(\ref{80}) can be derived by varying the Euclidean version
of action (\ref{76})
\begin{equation}
\label{81}
I_{D_2}^{Euc} = 2 \pi L T_2 
\int d\tau \left(\sqrt{(R^2 + \xi^2)(1 + \dot{R}^2)} - \frac{h}{2} R^2 \right).
\end{equation}

One can  show that eq.(\ref{80}) can be converted
 into the following first
order form
\begin{equation}
\label{82}
\sqrt{\frac{R^2 + \xi^2}{1 + \dot{R}^2}} - \frac{h}{2} R^2 = C,
\end{equation}
where $C$ is an integration constant. After some manipulations
 one can 
reexpress eq.(\ref{82}) in the following way:
\begin{equation}
\label{83}
\dot{R} = \frac{h}{2C + h R^2} 
\sqrt{(R_+^2 - R^2)(R^2 - R_-^2)}
\end{equation}
where 
\begin{eqnarray}
\label{84}
R_+^2 + R_-^2&=& \frac{4(1 - C h)}{h^2},    \\   \nonumber
R_+^2 R_-^2&=& \frac{4(C^2 - \xi^2)}{h^2}.
\end{eqnarray}
Comparing eq.(\ref{84}) with
corresponding  equations in refs.\cite{2,3},
 one can see  that
eq.(\ref{83}) is exactly the same as
 that of the  purely  electric case there if the electric
displacement $D$ there is identified with $\xi$.
 Thus  the general periodic instanton
solution of (\ref{84}) and its classical Euclidean action can be read 
off directly
from ref.\cite{3}. 

Here we consider only the  vacuum solution ($R_-=0, C=\xi$), which is 
determined by
\begin{equation}
\label{85}
\sqrt{R_+^2 - R^2} + \frac{2 \xi}{h R_+} \ln 
\frac{R_+ + \sqrt{R_+^2 - R^2}}{R} = -h (\tau_0 - \tau)
\end{equation}
where $R_+ = 2 \sqrt{1 - \xi h} / h$ and the corresponding Euclidean action is 
\begin{equation}
\label{86}
I_{cl} = N T_0 \int d\tau + L (\frac{4\pi}{3} R_+^3) \frac{h T_2}{2}.
\end{equation}
Here the first term  is
the  contribution of  $N$ D0-branes and the 
second term is the contribution of the  D2-brane.

Finally we show that  our formulation allows
 $D$ to be time-dependent but $N$ 
to be fixed. Using (\ref{71}) and (\ref{82}) $D$ is expressed 
in Euclidean space as 
\begin{equation}
\label{87}
D = \sqrt{\frac{1 + \dot{R}^2}{R^2 + \xi^2}} \xi 
= \frac{\xi}{C + \frac{h}{2} R^2}.
\end{equation}
Since $R$ is dependent on time, $D$ should also depend  on time. 
Using (\ref{71}) and (\ref{73}) $N$ can be generally expressed in Euclidean 
space as 
\begin{equation}
\label{88}
N = \frac{D}{2\pi \lambda} \int d\theta dz
\frac{R\sqrt{1 + R'^2 + \dot{R}^2}}
      {\sqrt{(1 + \dot{R}^2) (1 + \dot{R}^2 - D^2)}}.
\end{equation}
Thus if we consider $R=R(\tau)$,  $N$ reduces  to
\begin{equation}
\label{89}
N = \frac{L}{\lambda}
    \frac{DR}{\sqrt{1 + \dot{R}^2 - D^2}}.
\end{equation}
Inserting (\ref{87}) into (\ref{89}) we obtain
 $N = L \xi / \lambda$, which is our  
original  definition of $N$. 

One can also calculate the quantum-classical transition in this case
using the  periodic instanton or sphaleron solutions.
 The criterion for a first-order transition
can be read off directly from ref.\cite{3} as 
\begin{equation}
\label{90}
h \frac{N \lambda}{L}  <  \frac{1}{2}.
\end{equation}
Thus  the number of D0-branes as well as the RR-potential are involved in the 
criterion.

\section{Concluding remarks}

In the above we have considered D2-branes in the presence of spacetime
RR fields in the context of a model with world volume
cylindrical symmetry, and we have found  locally stable hemispherical
deformations of the brane,  the complementary hemispherical
configurations  being unstable.  We have also demonstrated
that these two configurations together comprise an
enveloping sphere representing a brane-antibrane pair
which in view of its associated Euclidean time
bounce configuration is unstable. This configuration
is analogous to the brane-antibrane configuration
constructed in ref. \cite{1} where the presence
of the bounce was anticipated.
The stability/instability  of the associated field configurations
 was investigated in detail in ref. \cite{13}, where it was
pointed out in particular that the brane-antibrane
configurations (constructed from the combination of a
stable brane and an unstable antibrane)
are again unstable.
We  have derived  explicitly the operator of
small fluctuations about such configurations
from which the  local stability or instability
of such configurations in the sense of minimising
the energy locally  may be
deduced.
We then calculated the corresponding transition
rate for the decay of such a locally
stable brane configuration through the hump of the RR  potential.
It might  be somewhat easier 
 to repeat the same steps in the case of a pure electric field
aligned with the RR field
in view of the  explicit symmetry of the
 action under the exchange of
time and space coordinates.
 More interesting  is the consideration of
the polar component of the electric field together with
 the  magnetic field.
In this case the action has
a nice symmetry as discussed in section 5.
 Also,  having  both electric
and magnetic fields in this case,
 one might expect the existence of both  strings as well as
D0-branes. Another interesting direction of
extension  is to consider the same phenomena
with D3-branes.
The  D3-brane is selfdual and physical quantities in different
regions of the energy and/or  coupling constant can have the same analytic
expression \cite{10}. But in all these  cases
 the states are unstable
because the strong RR field makes the potential unbounded
from  below. One might  try
to consider  D0 and spherical D2-branes
 applied to  a radially decreasing  RR field at infinity
to avoid this instability. The proper  principle, 
however, is to  somehow  take into
account  gravity, which might
 be a good candidate to  stabilize
the system as has been discussed, for instance in ref. \cite{8}.

\vspace{4cm}

\noindent
{\bf Acknowledgement}: S.T. acknowledges support by the Deutsche
Forschungsgemeinschaft (DFG).

\newpage

\begin{center}
{\large\bf Appendix A}
\end{center}

\begin{appendix}
\setcounter{equation}{0}
\renewcommand{\theequation}{A\arabic{equation}}
Here we summarise a few points in relation to  refs.\cite{4,10} in which
a spherical D2-brane was considered instead of
the cylindrical one  considered here.
In a simplified way  ref.\cite{10} has
\begin{eqnarray}
&& X^0=t,\;\;  X^1=r(t)\sin\theta\cos\phi, \;\; X^2=r(t)\sin\theta\sin\phi,
\nonumber\\
&& X^3=r(t)\cos\theta, \;\; others = const. (Dirichlet)
\label{1}
\end{eqnarray}
so that with $X^0=t$
$$
ds^2=-dt^2+dr^2+r^2(d\theta^2+\sin^2\theta d\phi^2) + \sum^9_{i=4}
(dX^i)^2.
$$
Now one takes as world volume coordinates $\xi_{\alpha}$ of the
$D2$-brane (denoted  by indices $\alpha,
\beta, \cdots$) the variables $t, \theta, \phi$.
Thus $r$, or the function $r(t)$, originally the third of
the three polar coordinates, acts as a scalar excitation
of the brane. 
Then
${ F}_{\alpha\beta}=\partial_{\alpha}
A_{\beta}-\partial_{\beta}A_{\alpha}$
and $ H=dA, H_{0123}=h$, 
and as  in \cite{4}
we take the background RR four-form field strength to be
$
H_{0123}=h=const\equiv h\epsilon_{123}.$
This is a field strength aligned with the 123 subspace of
spacetime, i.e. orthogonal to
the $4\cdots 9$ part.
The target space  metric
is thus changed from Minkowsky to ${\bf S}^2(r)\times{\bf R}^7$.
Then the Born-Infeld action integral becomes
\begin{equation}
I=\int d^3\xi{\cal L}(r, A_{\theta}, A_{\phi})
\label{2}
\end{equation}
where
\begin{eqnarray}
{\cal L}&=&-T_2
\bigg\{
 \bigg[-(-1+{\dot r}^2)
(r^4\sin^2\theta+4\pi^2 {F}^2_{\theta\phi})\nonumber\\
&&-4\pi^2r^2((\partial_0A_{\phi})^2+\sin^2\theta(\partial_0A_{\theta})^2)
\bigg]^{1/2}\nonumber\\
&&-\frac{hr^3\sin\theta}{3}
\bigg\}
\label{3}
\end{eqnarray}
The expression
$F_{\theta\phi}=\frac{N}{2}\sin\theta$
(apart from  its  normalisation)
is not a choice; rather it is dictated by the Euler-Lagrange
equations derived from ${\cal L}$ for  $F_{\theta\phi}$ in
the present case. Consider for simplicity the static case
and ignore  the Wess-Zumino contribution.
Set
$$
\frac{F_{\theta\phi}}
{ \sqrt{r^4\sin^2\theta +4\pi^2(\partial_{\theta}
A_{\phi}-\partial_{\phi}A_{\theta})^2}} \equiv D_s
$$
Then we obtain the equations 
$$
\frac{\partial D_s}{\partial\theta} = 0, \;\;
\frac{\partial D_s}{\partial\phi} = 0
$$
so that $D_s$ is independent of $\theta$ and $\phi$,
but, of course, depends om $r$. 
Now we can solve the former  equation for $F_{\theta\phi}$
and obtain
\begin{equation}
F_{\theta\phi} = \frac{D_sr^2\sin\theta}{\sqrt{1-D^2_s.4\pi^2}}
\label{4}
\end{equation}
This relation corresponds exactly to the relation one
obtains in the cylindrical case, there with $\sin\theta $
replaced by $R\sqrt{1+{R^{\prime}}^2}$.
Since
$$
2\pi N=\int d\phi d\theta F_{\theta\phi}
$$
the flux, obtained by integrating over the
closed surface of the 2-sphere  is
quantised. The number $N$ is identified with the number
of $D0$-branes.
From here on many of the considerations leading to 
the minimised energy parallel those in our considerations
above and therefore will not be given here. We cite
only the expression for the potential
\begin{eqnarray}
V_{D2}&=&4\pi T_2\bigg\{\sqrt{r^4+\pi^2N^2}-\frac{hr^3}{3}\bigg\}\nonumber\\
&=&\sqrt{(ST_2)^2+(NT_0)^2}-4\pi T_2\frac{hr^3}{3}.
\label{5}
\end{eqnarray}

\end{appendix}

\begin{center}
{\large\bf Appendix B}
\end{center}

\begin{appendix}
\setcounter{equation}{0}
\renewcommand{\theequation}{B\arabic{equation}}
We present  here the main steps involved in the determination
of the operator of small fluctuations since the  nontrivial procedure
can be useful in other analogous considerations.
Our starting point is the Hamiltonian
\begin{equation}
H=\frac{1}{2\pi g}\int dz\bigg(PR\sqrt{1+{R^{\prime}}^2}-\frac{h}{2}R^2\bigg)
\label{b1}
\end{equation}
where $P=1/\sqrt{1-D^2}, D=D(R)$, and for convenience
we set $p(D)\equiv PR\sqrt{1+{R^{\prime}}^2}
=\sqrt{R^2(1+{R^{\prime}}^2)+B^2}\equiv p(B)$. 
Thus, since $D=D(R)$, we first replace in the Hamiltonian $p(D)$ by
$p(B)$ and then perform the variation. Then after each variation
we can return to expressions in terms of $D$. 
Proceeding in this way the first variation yields
\begin{equation}
\frac{\delta H}{\delta R}=\frac{R(1+{R^{\prime}}^2)}{2\pi gp(B)}
-\frac{hR}{2\pi g},\;\;
\frac{\delta H}{\delta R^{\prime}}=\frac{R^2R^{\prime}}{2\pi g p(B)}.
\label{b2}
\end{equation}
From
\begin{equation}
\frac{\delta H}{\delta R}-\frac{d}{dz}\frac{\delta H}{\delta R^{\prime}}
=0
\label{b3}
\end{equation}
we obtain the equation of motion which when integrated and with
integration constant chosen equal to zero implies the relation
\begin{equation}
\frac{R}{P\sqrt{1+{R^{\prime}}^2}}-\frac{h}{2}R^2=0.
\label{b4}
\end{equation}
It is at the configuration given by this equation that the
second variation of $H$ is to be evaluated.
First  we obtain
\begin{eqnarray}
\frac{{\delta}^2H}{{\delta R}^2}
&=&\frac{1}{2\pi g}\bigg(\frac{1+{R^{\prime}}^2}{p(B)}-
\frac{R^2(1+{R^{\prime}}^2)^2}{{p(B)}^3} -h\bigg)\nonumber\\
&=&\frac{1}{2\pi g}\bigg[\frac{D^2}{P}\frac{\sqrt{1+{R^{\prime}}^2}}{R}
-h\bigg]
\label{b5}
\end{eqnarray}
and similarly (omitting now the intermediate step)
\begin{equation}
\frac{{\delta}^2H}{{\delta}{R^{\prime}}^2}
=\frac{1}{2\pi g}\frac{R(1+D^2{R^{\prime}}^2)}
{P(1+{R^{\prime}}^2)^{3/2}}, \;\;
\frac{{\delta}^2H}{{\delta R}{\delta {R}^{\prime}}}
=\frac{1}{2\pi g}\frac{R^{\prime}(D^2+1)}{P\sqrt{1+{R^{\prime}}^2}}
=\frac{{\delta}^2H}{{\delta {R}^{\prime}}{\delta R}}
\label{b6}
\end{equation}
These expressions are inserted into the second variation and give
\begin{eqnarray}
\delta^2H&=&\frac{1}{2\pi g}\frac{1}{2}\int dz\bigg[
\frac{{\delta}^2H}{{\delta}{R^{\prime}}^2}
(\delta R^{\prime})^2+
2\frac{{\delta}^2H}{{\delta R}{\delta {R}^{\prime}}}
\delta R\delta R^{\prime}
+
\frac{{\delta}^2H}{{\delta R}^2}(\delta R)^2\bigg]\nonumber\\
&=&\frac{1}{4\pi gP}\int dz [A]+\frac{D^2}{4\pi gP}\int dz [B]
\label{b7}
\end{eqnarray}
where
\begin{eqnarray}
A &=&\frac{R}{(1+{R^{\prime}}^2)^{3/2}}\bigg(\frac{d\delta R}{dz}\bigg)^2
+2\frac{R^{\prime}}{\sqrt{1+{R^{\prime}}^2}}\delta R\frac{d\delta R}{dz}
-hP(\delta R)^2,\nonumber\\
B &=&\frac{R{R^{\prime}}^2}{(1+{R^{\prime}}^2)^{3/2}}
\bigg(\frac{d\delta R}{dz}\bigg)^2
+2\frac{R^{\prime}}{\sqrt{1+{R^{\prime}}^2}}\delta R\frac{d\delta R}{dz}
+\frac{\sqrt{1+{R^{\prime}}^2}}{R}(\delta R)^2
\label{b8}
\end{eqnarray}
By writing
\begin{equation}
\frac{d}{dz}\delta R = R^{\prime}\bigg(\frac{d}{dz}\frac{\delta R}{R^{\prime}}
-\delta R\frac{d}{dz}\frac{1}{R^{\prime}}\bigg)
\label{b9}
\end{equation}
one can arrive after some manipulations at the expression
\begin{equation}
\bigg(\frac{d\delta R}{dz}\bigg)^2={R^{\prime}}^2\bigg(\frac{d}{dz}
\frac{\delta R}{R^{\prime}}\bigg)^2+R^{\prime\prime}\frac{d}{dz}\frac{
(\delta R)^2}{R^{\prime}}
\label{b10}
\end{equation}
We insert this expression into the expressions for $A$ and $B$.
Considering first the quantity $A$ and ignoring total
derivatives, we can rewrite this as
\begin{equation}
A=\frac{R{R^{\prime}}^2}{(1+{R^{\prime}}^2)^{3/2}}\bigg(\frac{d}{dz}\frac
{\delta R}{R^{\prime}}\bigg)^2+V(\delta R)^2,
\label{b11}
\end{equation}
where
\begin{equation}
V =-\frac{1}{R^{\prime}}\frac{d}{dz}
\frac{RR^{\prime\prime}}{(1+{R^{\prime}}^2)^{3/2}}
-\frac{d}{dz}\frac{R^{\prime}}{\sqrt{1+{R^{\prime}}^2}}-hP
\label{b12}
\end{equation}
One can show that for the solutions of eq. (\ref{b4}) $ V=0$.
 Considering
now  the quantity $B$ and ignoring total derivatives, we can rewrite
this as
\begin{equation}
B=\frac{R{R^{\prime}}^4}{(1+{R^{\prime}}^2)^{3/2}}\bigg(\frac{d}{dz}\frac
{\delta R}{R^{\prime}}\bigg)^2+ U(\delta R)^2,
\label{b13}
\end{equation}
where
\begin{equation}
U =-\frac{1}{R^{\prime}}\frac{d}{dz}
\frac{R{R^{\prime}}^2R^{\prime\prime}}{(1+{R^{\prime}}^2)^{3/2}}
-\frac{d}{dz}\frac{R^{\prime}}{\sqrt{1+{R^{\prime}}^2}}+\frac
{\sqrt{1+{R^{\prime}}^2}}{R}.
\label{b14}
\end{equation}
One can show that for the solutions of eq. (\ref{b4}) $ U=0$.
Thus finally we are left with (replacing $dz$ by $dR/R^{\prime}$)
\begin{equation}
\delta^2H=\frac{1}{4\pi gP}\int \frac{dR}{R^{\prime}}
\frac{R{R^{\prime}}^2(1+D^2{R^{\prime}}^2)}
{(1+{R^{\prime}}^2)^{3/2}}\bigg(\frac{d}{dz}\frac
{\delta R}{R^{\prime}}\bigg)^2.
\label{b15}
\end{equation}
Thus this remaining term is positive definite
for positive derivative $R^{\prime}$,
and negative for negative $R^{\prime}$.
Thus for solutions $R$ with $R^{\prime}$ positive
the energy is minimised and for those with
$R^{\prime}$ negative it is maximised.  We can now write
the second variation
\begin{equation}
\delta^2H=\frac{1}{2}\int dz\delta R{\hat M}\delta R
\label{b16}
\end{equation}
where the fluctuation operator ${\hat M}$ is
\begin{equation}
{\hat M}=-\frac{1}{2\pi gP}\frac{1}{R^{\prime}}\frac{d}{dz}
\frac{R{R^{\prime}}^2(1+D^2{R^{\prime}}^2)}
{(1+{R^{\prime}}^2)^{3/2}}\frac{d}{dz}\frac{1}{R^{\prime}}
\label{b17}
\end{equation}

\end{appendix}

\begin{center}
{\large\bf Appendix C}
\end{center}

\begin{appendix}
\setcounter{equation}{0}
\renewcommand{\theequation}{C\arabic{equation}}

The action $I$ is given by eq.(\ref{7}) with the
Lagrangian (\ref{8}) and in this the magnetic field
$B$ is  given by eq. (\ref{11}). Then we make a Wick
rotation by setting
 $t=i\tau$. The resulting Euclidean action is
\begin{equation}
I_E=\int d\tau dzd\theta {\cal L_E}
\label{a4}
\end{equation}
 where (dots now refer to Euclidean time)
\begin{equation}
{\cal L_E} = \frac{1}{4\pi^2g}\bigg\{-
\sqrt{R^2(1+{\dot R}^2+{R^{\prime}}^2)
+B^2(1+{\dot R}^2)}+\frac{h}{2}R^2\bigg\}.
\label{a5}
\end{equation}
with magnetic field $B$ given by
\begin{equation}
B^2=\frac{D^2 R^2}{1+{\dot R}^2}\frac{1+{R^{\prime}}^2+{\dot R}^2}{1+{\dot R}^2-D^2}
\label{a6}
\end{equation}
The induced metric is
\begin{eqnarray}
g_{\alpha \beta}&=&\left(\begin{array}{ccc}
1+{\dot R}^2 & {\dot R}R^{\prime} & 0\\
R^{\prime}{\dot R} & 1+{R^{\prime}}^2 & 0 \\
0 & 0 & R^2
\end{array}\right)
\label{a7}
\end{eqnarray}
and the field tensor (cf. eq. (\ref{6}))
\begin{eqnarray}
2\pi F_{\alpha \beta}&=&\left(\begin{array}{ccc}
0 & 0 & 0\\
0 & 0 & B \\
0 & -B & 0
\end{array}\right)
\label{a8}
\end{eqnarray}
Now we make the  right angle rotation to
\begin{equation}
z\rightarrow \tilde z=\tau, \qquad  \tau\rightarrow \tilde \tau=-z
\label{a9}
\end{equation}
so that (omitting tildes on $R$ for simplicity)
\begin{equation}
R\rightarrow R, \qquad R^{\prime}\rightarrow
{-\dot R}, \qquad {\dot R}\rightarrow R^{\prime}.
\label{a10}
\end{equation}
The magnetic field $\tilde B$ becomes
\begin{equation}
{\tilde B}^2=\frac{D^2 R^2}
{1+{R^{\prime}}^2}\frac{1+{R^{\prime}}^2+{\dot R}^2}{1+{R^{\prime}}^2-D^2}
\label{a11}
\end{equation}
and the induced metric $\tilde g$
\begin{eqnarray}
\tilde g_{\alpha \beta}&=&\left(\begin{array}{ccc}
1+{\dot R}^2 & -{\dot R}R^{\prime} & 0\\
-R^{\prime}{\dot R} & 1+{R^{\prime}}^2 & 0 \\
0 & 0 & R^2
\end{array}\right)
\label{a12}
\end{eqnarray}
and the field tensor
\begin{eqnarray}
\tilde F_{\alpha \beta}&=&\left(\begin{array}{ccc}
0 & 0 & \tilde B\\
0 & 0 & 0 \\
-\tilde B & 0  & 0
\end{array}\right).
\label{a13}
\end{eqnarray}
These expressions together yield the Lagrangian
\begin{equation}
{\tilde {\cal L}_E} = \frac{1}{4\pi^2g}\bigg\{-
\sqrt{R^2(1+{\dot R}^2+{R^{\prime}}^2)
+{\tilde B}^2(1+{R^{\prime}}^2)}+\frac{h}{2}R^2\bigg\}.
\label{a14}
\end{equation}
and the action
\begin{equation}
\tilde I_E=\int d \tilde \tau d \tilde zd\theta {\tilde {\cal L}_E}
\label{a15}
\end{equation}
Since the Lagrangian is a Lorentz scalar the result could also
have been written down directly from (\ref{a5}).
For the instanton solution $R^{\prime}=0$, and the Lagrangian assumes
 the following form which demonstrates its
equivalence  with that of the static solution:
\begin{equation}
{\tilde {\cal L}_E} = \frac{1}{4\pi^2g}\bigg\{-
PR\sqrt{1+{\dot R}^2}+\frac{h}{2}R^2\bigg\}.
\label{a16}
\end{equation}
\vspace{0.5cm}
\noindent

\end{appendix}

\newpage
\vspace{0.2cm}
\centerline{Fig. 1 $\;\;$ The two circular shells
 and the enveloping sphere}
\vspace{0.2cm}

\newpage
\epsfysize=10cm \epsfbox{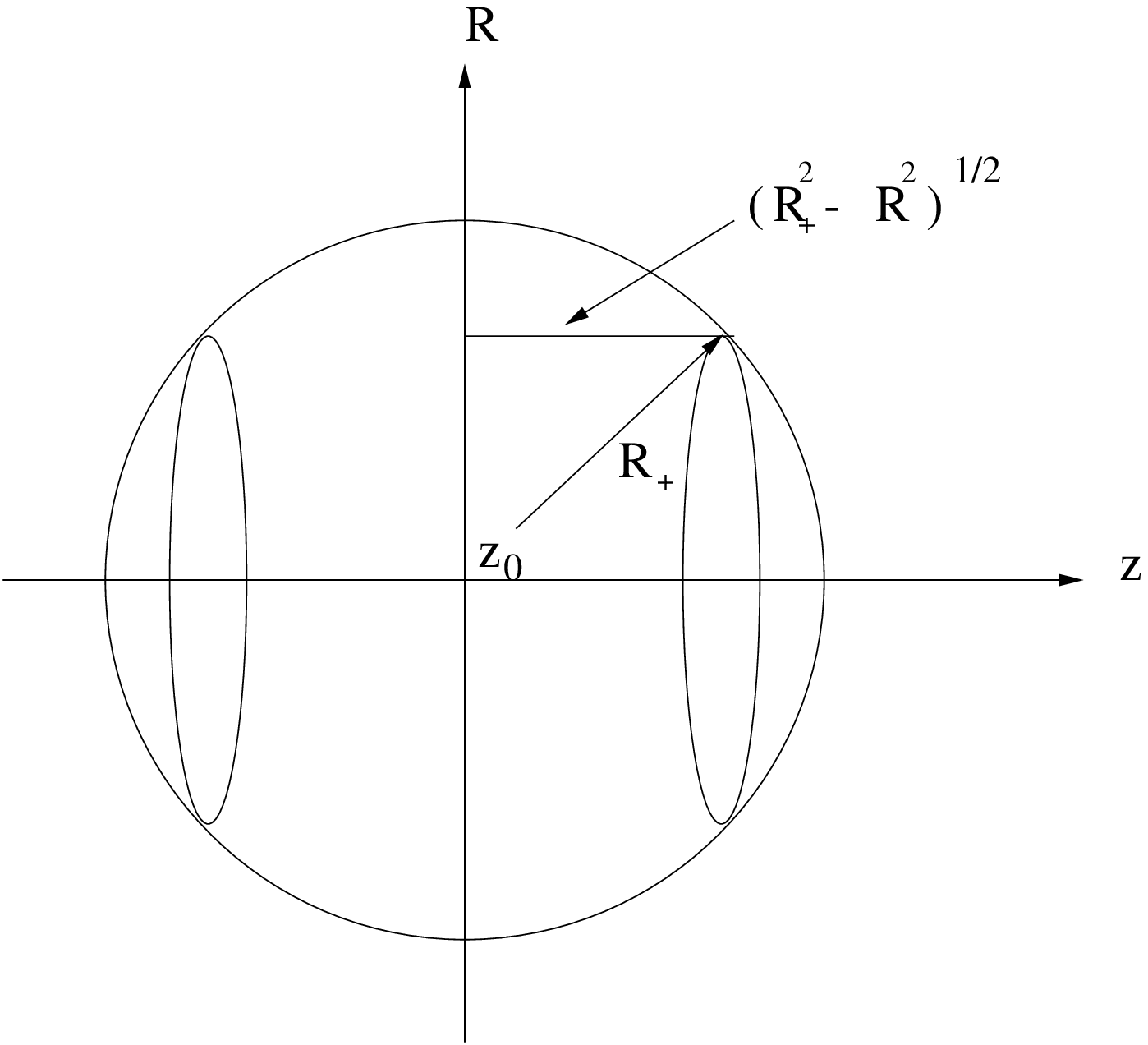}

\end{document}